\documentclass{article}
\usepackage{arxiv/PRIMEarxiv}

\usepackage{tikz}
\usepackage{booktabs}
\usepackage{mathtools}
\DeclarePairedDelimiter\ceil{\lceil}{\rceil}
\DeclarePairedDelimiter\floor{\lfloor}{\rfloor}
\usepackage{amssymb, amsmath, amsthm, color, tikz,mathrsfs}
\usepackage{pdfpages}
\usepackage{hyperref}
\usepackage{textcomp}
\usepackage{color}
\usepackage{xcolor}
\usepackage{listings}
\usepackage{algorithm}
\usepackage{algpseudocode}
\usepackage{multirow}
\usepackage{subcaption}

\newcommand{\vanish}[1]{}

\AtBeginDocument{%
  
}

\algrenewcommand\algorithmicrequire{\textbf{Input:}}
\algrenewcommand\algorithmicensure{\textbf{Output:}}

\title{Optimization Perspectives on Shellsort}

\author{
  Oscar Skean \\
  Department of Computer Science\\
  University of Kentucky \\
  \texttt{oscar.skean@uky.edu} \\
   \And
  Richard Ehrenborg \\
  Department of Mathematics\\
  University of Kentucky \\
  \And
  Jerzy W. Jaromczyk \\
  Department of Computer Science\\
  University of Kentucky \\
}
\begin{document}

\maketitle

\begin{abstract}
Shellsort is a sorting method that is attractive due to its simplicity, yet it takes effort to analyze its efficiency. The heart of the algorithm is the gap sequence chosen a priori and used during sorting. The selection of this gap sequence affects the efficiency of Shellsort, and thus drives both its theoretical and experimental analysis. We contribute to Shellsort by identifying efficient gap sequences based on new parameterized functions. Specifically, a parameter grid-search identifies optimal parameters for different input sizes for sorting by observing minimal overhead in three categories: number of comparisons, number of exchanges, and running time. We report that our method finds sequences that outperform state-of-the-art gap sequences concerning the number of comparisons for chosen small array sizes. Additionally, our function-based sequences outperform the running time of the Tokuda sequences for chosen large array sizes. However, no substantial improvements were observed when minimizing the number of exchanges.
\end{abstract}

\section{Introduction}

The Shellsort algorithm is a sorting method that was among the first to be discovered. Published in 1959~\cite{shell1959high}, it saw early interest due to its low memory requirements and simple implementation. Despite this, its analysis is difficult and remains incomplete. The algorithm has found practical use today in memory-constrained environments, embedded systems, and the bzip2 compressor. Recently it has also found use in data-oblivious sorting~\cite{goodrich2011randomized} and in fully homomorphic encryption~\cite{lee2021analysis}.

Shellsort is an in-place comparison sort and can be viewed as a generalization of insertion sort. For a data array $A$ of size $N$, Shellsort operates using a predetermined gap sequence $1 = k_1 < \cdots < k_m < N$. The algorithm performs $m$ passes over $A$: starting with the largest $k_m$ and ending with $k_1$. During pass $j$, for a given gap $k_{m-j}$, insertion sort occurs for the $k_{m-j}$ subarrays consisting of the data elements $A(i), A(i+k_{m-j}), A(i+2 \cdot k_{m-j}), \ldots$ for $i = 0, \ldots, k_{m-j}-1$. We say a {\em $k$-inversion} is a pair $(i,i+k)$ such that the inequality $A(i) > A(i+k)$ holds. After pass $j$, all $k_{m-j}$-inversions that were originally in $A$ have been solved. We say an array with no $k$-inversions is $k$-sorted. Note that the final pass with $k_1$ is equivalent to insertion sort and is necessary to guarantee sortedness. Therefore, the purpose of the gap sequence is to presort $A$ as much as possible before the expensive final insertion sort pass.

The main results of this paper are:

\begin{enumerate}
\item{New efficient Shellsort sequences derived from experimentally optimizing sequence-generating functions. For prescribed array sizes, these sequences outperform well-known efficient sequences (e.g. Tokuda, Ciura) with respect to the number of comparisons. These sequences also outperform the running time of the Tokuda sequence, making them the fastest function-based sequence on the tested array sizes. } 

\item{We demonstrate results of experimental analysis comparing our proposed approaches with well-known sequences by measuring the number of comparisons, exchange operations, and running time needed to sort randomized permutations.}
\end{enumerate}

Traditionally, improvements for Shellsort have come from finding gap sequences with theoretical properties. We discuss some particularly important sequences in Section \ref{Background}. Then in Section \ref{templates}, we introduce parameterized sequence-generating functions that generate a Shellsort sequence. The parameters are then optimized in a grid-search finding the best possible sequence that can be produced from that function for a chosen array size. In Section \ref{templateexp} we discuss our experimental methodology to compare the performance of the optimized template sequences to the baseline sequences mentioned in Section \ref{Background}.

\section{Background}
\label{Background}

The selection of a good gap sequence is critical to the performance of Shellsort. There has been a plethora of work focused on selecting good sequences~\cite{ciura2001best, sedgewick1996analysis, tokuda1992improved}. Some of the earliest proposed sequences were based on powers of 2~\cite{hibbard1963empirical, shell1959high}. Then Pratt showed that the sequence of $2^p 3^q$ obtains a number of inversions that is $\Theta(N log^2 N)$ in the worst case~\cite{pratt1972shellsort}. We call this sequence Pratt-23 in Table \ref{table:masterlist}. This sequence still has the best known asymptotic time complexity for any Shellsort sequence. However, it has a very large constant factor which spurred the development of new sequences. 

The proof technique used to show the time complexity of Pratt-23 was based on counting the inversions of a sequence that has already been 2-sorted and 3-sorted. A natural extension of this is to apply the Frobenius problem to place bounds on what has already been sorted in prior passes. A typical formulation of the Frobenius problem is as follows: Suppose that you have $k$ coins
of denominations $u_1, u_2, \ldots, u_k$. What is the largest value which cannot be made with a nonnegative linear combination of these coins? This largest value is known as the Frobenius number~\cite{selmer1977linear}. In the context of Shellsort, the coins can be equated to gap size and the Frobenius number can be equated to the largest remaining inversion after sorting with the gaps. Using the Frobenius problem, several sequences were proposed that had a lower constant factor than Pratt-23 despite having a worse time-complexity~\cite{incerpi1985improved, selmer1987shellsort}.

Following those sequences, the focus of gap sequence selection shifted from finding theoretically good sequences to finding experimentally good ones. For example, one property that was observed was that a geometric sequence with a growth of $2.25$ often performed well in practice. This observation was the basis of the Tokuda sequence~\cite{tokuda1992improved}. See Table~\ref{table:masterlist} for a functional form. To the best of our knowledge, this remains the most competitive function in the literature. 

The next improvement came from Ciura in 2002~\cite{ciura2001best}. The Ciura sequence disregarded the idea of a function-based sequence, and instead searched for the best set of gap elements themselves. Ciura found the best sequence for array sizes of 128 and 1000, as well as a sequence that was conjectured to perform better for much larger array sizes. We call these Ciura-128, Ciura-1000, and Ciura-Large in Table \ref{table:masterlist}.

The Ciura sequences also marked a transition in how Shellsort performance was measured. Previously, most works counted the number of exchanges used by the algorithm~\cite{pratt1972shellsort}. This partly because some proof techniques relied on counting the number of inversions. Ciura instead focused on optimizing the number of comparisons, which was found to be more directly related to the computation time of Shellsort. A comparison is defined as checking if a pair of array elements are inverted. An exchange is typically defined as the variable swap used to fix an inversion. Minimizing the number of comparisons is especially beneficial when comparisons are expensive to make, such as when sorting large satellite data. Similarly, minimizing exchanges is beneficial in memory-constrained systems. In this work, we make clear distinctions in Section~\ref{templateexp}  about what measurement we're optimizing for. For a full treatment of the history of gap sequences, we point the reader to~\cite{sedgewick1996analysis}.

\begin{table}
\centering
\resizebox{\columnwidth}{!}{%
\begin{tabular}[t]{ccccl}
\toprule
     \textbf{Sequence Name} & \textbf{Function} & \textbf{Optimized for N} &  \textbf{Parameters}  & \textbf{Initial Terms} \\ 
\midrule
     Ciura~\cite{ciura2001best}  & - & 128   & - & 1 4 9 24 85 126    \\
      & - & 1000 & - & 1 4 10 23 57 156 409 995 \\
      & - & Large & - & 1 4 10 23 57 132 301 701 1750\\
  \midrule
    Tokuda~\cite{tokuda1992improved} & $\ceil{\frac{(9/4)^k - 1}{(9/4) - 1}}$  & -  & - &  1 4 9 20 46 103 233 525 $\ldots$ \\
  \midrule
    Ours A & $\floor{(a^{\floor{\frac{i}{b}}} \cdot c^{\floor{\frac{i}{d}}})^f + e}$ & 128 (Comp) & Table \ref{table:params} & 1 4 9 24 85 150 $\ldots$\\
    & & 1000 (Comp) & Table \ref{table:params} & 1 4 10 23 57 153 400 $\ldots$\\
    & & 1000 (Time) & Table \ref{table:params} & 1 3 7 16 33 85 179 472 $\ldots$\\
  \midrule
   Ours B & $\floor{(a \cdot b ^ {\floor{\frac{i}{c}}}) + d}$ & 10000 (Comp) & Table \ref{table:params} & 1 4 10 27 72 187 488 $\ldots$ \\
  \midrule
   Pratt-23~\cite{pratt1972shellsort}  & Ordered $2^p \cdot 3^q$ & -& - & 1 2 3 4 6 8 9 $\ldots$ \\
   \midrule
  Pratt-25 & Ordered $2^p \cdot 5^q$ & - & - & 1 2 4 5 8 10 15 16 $\ldots$ \\
  
  Pratt-34 & Ordered $3^p \cdot 4^q$ & & - & 1 3 4 9 12 16 24  $\ldots$ \\
  \bottomrule                          
\end{tabular}
}
\caption{Gap sequences that are compared during experiments}
\label{table:masterlist}
\end{table}

\begin{table}
\centering
\begin{tabular}{|l|l|l|l|l|l|l|}
\hline
Template & $a$      & $b$      & $c$      & $d$      & $e$ & $f$      \\ \hline
Ours-A128-Comp    & 2.6321   & 1.6841  & 2.1570  & 0.7360  & 3 & 0.7630  \\
Ours-A1000-Comp   & 3.5789 & 2.6316 & 3.8158 & 2.1579 & 3 & 0.7632 \\
Ours-A1000-Time   & 2.75 &  2.75 & 3.7142 & 2.4286 & 2 &         0.7429 \\
Ours-B10000-Comp   & 4.0816 & 8.5714 & 2.2449 & 0 & - & - \\
\hline
\end{tabular}
\caption{Optimized parameters for template functions}
\label{table:params}
\end{table}

\section{Parameterized Template Functions}
\label{templates}
The approach of directly optimizing the gap sequence, as in~\cite{ciura2001best}, grows in computational cost very quickly as $N$ increases. This growth is due the fact that as $N$ increases, both the expected number of sequence elements and their possible range of values increases. To help alleviate this, the authors of~\cite{ciura2001best} found a suitable sequence prefix and optimize the extension of it by a few values. However, at very large $N$ even finding this prefix would be very costly. 

Here, we formulate the problem of finding the optimal sequence as optimizing the parameters of a pre-defined sequence-generating function. Guided by principles that we have seen in sequences that perform well, we define two functions as follows.
\vspace*{-2mm}
\begin{align} \label{eq:formatA}
k_A(i) & = \floor{(a^{\floor{\frac{i}{b}}} \cdot c^{\floor{\frac{i}{d}}})^f + e}\\
\label{eq:formatB}
k_B(i) & = \floor{(a \cdot b ^ {\floor{\frac{i}{c}}}) + d}
\end{align}
We refer to~\eqref{eq:formatA} as Ours-A and~\eqref{eq:formatB} as Ours-B in Table~\ref{table:masterlist}. Both formats contain the floor function in exponents. We found this to allow the function to express a more "chaotic" sequence, which helped improve performance. The unique characteristic of Ours-A is the parameter $f$, an exponent that helps regulate growth. The parameter $a$ of Ours-B was designed to have a similar purpose, albeit via multiplication. Ours-B contains fewer parameters which allows for quicker optimization. For conciseness, we only optimize Ours-A for array sizes 128 and 1000, and Ours-B for 10000. Furthermore, we denote as Ours-A1000-Comp if optimizing Format A for number of comparisons on arrays of size 1000. In the following section, we discuss the experimental procedure for optimizing (\ref{eq:formatA}) and (\ref{eq:formatB}), as well as for comparing them to other baseline sequences.

\section{Experimental Procedure}
\label{templateexp}
Because the function is highly non-convex, it is difficult to utilize efficient techniques such as gradient descent. Instead, we employ a grid-search approach.  One benefit of using a function grid-search, as opposed to direct sequence optimization, is that the size of the search space has no relation to $N$. It depends only on the granularity and bounds of the search. This is in constrast to the methodology used by Ciura, in which the number of tested sequences grows with $N$ \cite{ciura2001best}.

For Ours-A, we define the grid for parameters $a,b,c,d,f$ as 20 linearly spaced values between 0.5 and 5. We also allow $e$ to be an integer value between 0 and 10, including both endpoints. Because Ours-B has fewer parameters, we can take a more fine-grained approach. For parameters $a,b,c$, we test 50 linearly spaced values from 0 to 10. For parameter $d$, we constrain it to be the same as $e$ in Ours-A.

The data array that we test on contains $N$ distinct values $1$ through $N$, and we shuffle it with the Fischer-Yates shuffle. For each set of parameters in the grid-space, we compute the mean cost over 1000 iterations. We then take the set of parameters producing the lowest mean cost as optimal. This cost can be defined as number of comparisons, number of exchanges, or time.  

Because Ciura sequences are optimized for specific array sizes with no means of extending them, it would be inappropriate to directly compare them with any function-based sequence at large array sizes. One method of extending a Ciura sequence is by starting a geometric series on its last term with a ratio of 2.25. We adopt this method of extension when measuring the performance of Ciura sequences. 

\newpage

\subsection{Filtering}
There are two techniques we employ to reduce our grid-search space.

First, we notice that different sets of parameters could produce the same sequence. For example, for the template function Ours-A, the ordering of $(a,b)$ and $(c,d)$ do not matter. We precalculate all of the sequences produced in the grid and only experiment on the unique ones. For example, for $N = 10000$, this reduces the grid search space from over 1.5 million sets of parameters to about 1 million.

Second, we use sequential analysis to act as a low-pass filter for screening out obviously poor sequences. This statistical approach was first applied to Shellsort in~\cite{ciura2001best}. Given bounds for the mean and an upper bound for the variance, sequential analysis is able to tell in just a few repetitions whether or not a sample mean falls below the mean bounds with a certain confidence. Sequential analysis allows us to quickly accept good gap sequences that have a low mean number of comparisons. Any sequence that is accepted by the filter is then run for the full 1000 iterations to obtain a more accurate estimate of the mean. We adopt the same setup as in~\cite{ciura2001best}.

\begin{table}
\centering
\resizebox{\columnwidth}{!}{%
\begin{tabular}{|p{3cm}|l|l|l|l|l|l|}
\hline
  \multirow{2}{5cm}{Sequence} &   
  \multicolumn{2}{c|}{\textbf{N=20}} &
  \multicolumn{2}{c|}{\textbf{N=128}} & \multicolumn{2}{c|}{\textbf{N=200}}
\\
        \cline{2-7}
            & $\mu_{CO}$ & $\mu_{EX}$ & $\mu_{CO}$ & $\mu_{EX}$ & $\mu_{CO}$ & $\mu_{EX}$\\
        \hline
 Ours-A128-Comp &  \: 76 ± 6 &  38 ± 6  &  \: \textbf{998 ± 33} & 531 ± 33  & 1786 ± 46 & 948 ± 48  \\
Ours-A1000-Comp &  \: 76 ± 6 &  39 ± 7 & 1004 ± 32 &  516 ± 31   & 1787 ± 44 & 919 ± 45\\
Ours-A1000-Time & \:  79 ± 5 & 39 ± 7 & 1035 ± 26 & 468 ± 27 & 1832 ± 38 & 846 ± 39 \\
         Ours-B10000-Comp &  \: 76 ± 7 &  33 ± 5 & 1096 ± 52 & 535 ± 36 & \textbf{1775 ± 49} & 960 ± 49 \\
          Ciura-128 &  \: 76 ± 6 &  37 ± 6 &  \: \textbf{998 ± 32} & 531 ± 33 & 1800 ± 46 &  970 ± 49 \\
         Ciura-1000 &  \: 76 ± 7 &  39 ± 7 & 1006 ± 31 & 519 ± 34 & 1787 ± 45 & 920 ± 44 \\
         Ciura-Long &  \: 76 ± 7 &  39 ± 7  & 1004 ± 32 &  516 ± 32 & 1794 ± 44 & 907 ± 42 \\
            Tokuda &  \: 76 ± 6 &  37 ± 6 & 1020 ± 28 &  490 ± 28 & 1808 ± 42 &  891 ± 43 \\
            Pratt-25 &  111 ± 4 &   27 ± 4   & 1732 ± 16 & 345 ± 17  & 3207 ± 21 & 610 ± 24 \\
             Pratt-23 &  136 ± 3 &  25 ± 4 & 2209 ± 13 & \textbf{333 ± 15}  & 4095 ± 19 &  \textbf{589 ± 21}  \\
            Pratt-34 &  \: 95 ± 4 & 29 ± 4 & 1424 ± 16 & 374 ± 19  & 2593 ± 25 &  660 ± 26 \\

\hline
\end{tabular}
}
\caption{Number of operations to sort small arrays averaged over 1000 random array permutations. $\mu_{CO}$ denotes the number of comparisons, $\mu_{EX}$ denotes the number of exchanges.}\label{Tab:smalltable}
\end{table}

\begin{table}

\centering
\resizebox{\columnwidth}{!}{%
\begin{tabular}{|p{3cm}|l|l|l|l|l|l|}
\hline
  \multirow{2}{5cm}{Sequence} &   
  \multicolumn{2}{c|}{\textbf{N=1000}} &
  \multicolumn{2}{c|}{\textbf{N=2000}} &
  \multicolumn{2}{c|}{\textbf{N=5000}}
\\
        \cline{2-7}
            & $\mu_{CO}$ & $\mu_{EX}$ & $\mu_{CO}$ & $\mu_{EX}$ & $\mu_{CO}$ & $\mu_{EX}$  \\
        \hline
   Ours-A128-Comp & 13250 ± 203 & 7847 ± 199  & 30530 ± 378 & 18611 ± 384 &  \:  91122 ± 973 & \: 57728 ± 904 \\
  Ours-A1000-Comp & 12941 ± 167 & 7004 ± 155 & 29596 ± 293 & 16234 ± 282 &  \:  86821 ± 768 & \: 50349 ± 770 \\
  Ours-A1000-Time &  13193 ± 144 & 6461 ± 146 & 30120 ± 263 & 14913 ± 257 & \: 87455 ± 548 & \: 44305 ± 552  \\
 Ours-B10000-Comp & 12980 ± 186 & 7245 ± 177 & 29643 ± 305 & 17241 ± 325 & \:  86514 ± 617 &   \: 57388 ± 817 \\
      Ciura-128 & 13300 ± 166 & 7003 ± 168 & 30359 ± 318 & 15987 ± 310 &  \: 88193 ± 629 &   \: 46689 ± 627  \\
     Ciura-1000 & \textbf{12918 ± 161} & 7002 ± 155 & \textbf{29534 ± 282 }& 16138 ± 274 & \:  86641 ± 757 &   \: 47852 ± 751 \\
     Ciura-Long & 13035 ± 142 &  6701 ± 149 & 29567 ± 246 & 15427 ± 261 & \:  \textbf{86232 ± 502 }&   \: 45347 ± 496  \\
        Tokuda & 13116 ± 143 & 6556 ± 142   & 29888 ± 241 & 14952 ± 228 & \:  86838 ± 454 &   \: 44116 ± 472 \\
        Pratt-25 &  26211 ± 68 & 4318 ± 72 & 62722 ± 122 & \: 9755 ± 131 & 194196 ± 263 &   \: \textbf{28195 ± 278}

        \\
         Pratt-23 &  34380 ± 64 &   \textbf{4253 ± 69} & 82785 ± 106 &   \: \textbf{9669 ± 116} & 259088 ± 242 &   \: 28354 ± 257  \\
        Pratt-34 &  20974 ± 89 & 4671 ± 87 & 50038 ± 153 & 10543 ± 160 & 154298 ± 372 &   \: 30448 ± 372  \\
\hline
\end{tabular}%
}
\caption{Number of operations to sort medium-sized arrays averaged over 1000 random array permutations}\label{Tab:medtable}
\end{table}

\begin{table}
\centering
\scalebox{0.99}{%
\begin{tabular}{|p{3cm}|l|l|l|l|}
\hline
  \multirow{2}{5cm}{Sequence} &   
  \multicolumn{2}{c|}{\textbf{N=10000}}
\\
        \cline{2-3}
            & $\mu_{CO}$ & $\mu_{EX}$\\
        \hline
  Ours-A128-Comp & 206356 ± 1796 & 132351 ± 1797 \\
 Ours-A1000-Comp & 196336 ± 1707 & 119012 ± 1710 \\
 Ours-A1000-Time &  194052 ± 879 &  \:  98952 ± 883 \\
 Ours-B10000-Comp &  192029 ± 992 & 209292 ± 1293 \\
       Ciura-128 & 195256 ± 1106 & 105544 ± 1109 \\
      Ciura-1000 & 193778 ± 1895 & 111338 ± 1897 \\
      Ciura-Long &  \textbf{191435 ± 892} &  101680 ± 897 \\
          Tokuda &  192574 ± 795 &  \:  98071 ± 796 \\
        Pratt-25 &  450131 ± 516 &  \:  \textbf{62191 ± 526} \\
        Pratt-23 &  604502 ± 451 & \:   66923 ± 725\\
        Pratt-34 &  355382 ± 723 & \:   63272 ± 462  \\
\hline
\end{tabular}%
}
\caption{Number of operations to sort an array size of 10000 averaged over 1000 random array permutations}\label{Tab:bigtable}
\end{table}

\begin{table}
\centering
\begin{tabular}{|l|l|}

\hline
            Sequence &       Running Time (ms) \\
\hline
  Ours-A128-Comp & 3.15 ± 0.08\\
 Ours-A1000-Comp & 3.02 ± 0.06\\
 Ours-A1000-Time & \textbf{3.01 ± 0.06}\\
Ours-B10000-Comp & 3.04 ± 0.07\\
       Ciura-128 & 3.07 ± 0.06\\
      Ciura-1000 & \textbf{3.01 ± 0.06}\\
      Ciura-Long & 3.04 ± 0.07\\
          Tokuda & 3.06 ± 0.08\\
        Pratt-25 & 5.00 ± 0.09\\
        Pratt-23 & 6.35 ± 0.11\\
        Pratt-34 & 4.17 ± 0.08 \\
\bottomrule
\end{tabular}

\caption{Time to sort an array size of 1000 averaged over 1000 random array permutations}\label{Tab:timetable}
\end{table}

The Ciura sequences were optimized with respect to the number of comparisons, and because they are well-known to be some of the most practical sequences, we optimize our template functions with respect to the number of comparisons as well.

\begin{figure}
\centering
    \begin{subfigure}{.48\textwidth}
          \centering
        \includegraphics[width=\textwidth]{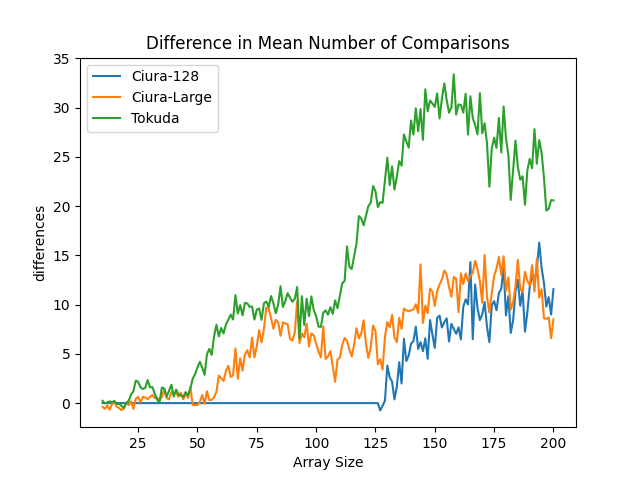}
        \label{fig:meandiffs}
    \end{subfigure}%
    \begin{subfigure}{.48\textwidth}
          \centering
        \includegraphics[width=\textwidth]{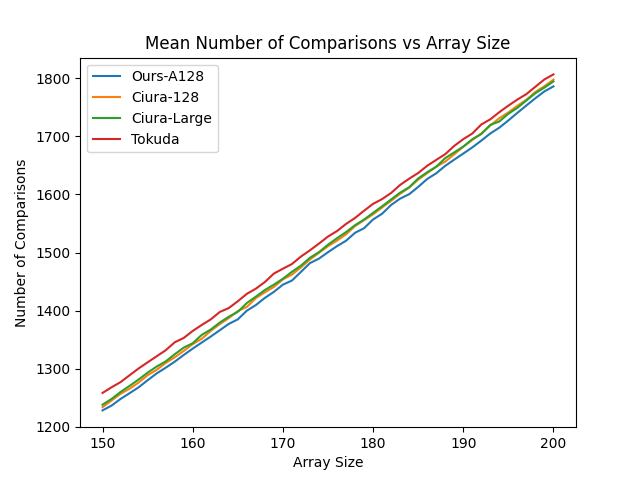}
        \label{fig:comparisons}
    \end{subfigure}
\caption{(Left) For varying array sizes, shows the difference in number of comparisons between baseline sequences and Ours-A128. A positive value means Ours-A128 uses fewer comparisons. (Right) Number of comparisons for varying array sizes larger than what Ours-A128 was optimized for.}\label{fig:graphfig}
\end{figure}

\subsection{Hardware}
The experiments were performed on a Ubuntu machine with an 8-core Intel Xeon W-3225. Experiments counting number of comparisons and exchanges were multithreaded. Experiments involving measurement of time were done single threaded, as any multithreaded applications could cause discrepancies in time measurement. All code was written in Python.

\subsection{Results}

The best parameters that we found for Ours-A and Ours-B are in Table \ref{table:params}. Additionally, the first few terms of the sequences are shown in Table \ref{table:masterlist}. 

For array size 200, we have found that Ours-B10000-Comp outperforms all other tested sequences in terms of number of comparisons, as shown in Table \ref{Tab:smalltable}. Figure \ref{fig:graphfig} is a graphical aid to this table. These graphs show that sequences generated by template functions can still perform well for array sizes larger, but not significantly so, than what they were optimized for. Furthermore, we found that both Ours-A128-Comp and Ours-A1000-Comp met the number of comparisons of, but did not surpass, the Ciura and Tokuda sequences. It's interesting to note that several of the initial terms are equivalent between Ciura-128 and Ours-A128-Comp.

We also test medium and large arrays, with results shown in Tables \ref{Tab:medtable} and \ref{Tab:bigtable}. For a graphical representation of Table \ref{Tab:bigtable}, see Figure 1 in the Appendix. Our new sequences approach the performance of Ciura sequences without surpassing them. However, Ours-B10000 still surpasses the Tokuda sequence for all array sizes that we have tested here. Recall that the Tokuda sequence is currently the best known sequence to be generated from a function. Therefore, we have shown a new function-based sequence that outperforms other function-based sequences in terms of the number of comparisons. As mentioned previously, this is particularly useful if comparisons are a dominant operation such as when sorting large satellite data.

On the other hand, our experiments with optimizing sequences to minimize running time are shown in Table \ref{Tab:timetable}. The relevant sequence, Ours-A1000-Time, takes a similar running time to the Ciura-1000 sequence. Both are faster than any other tested sequence. The sequence Ours-A1000-Time is particularly interesting because its first few elements (1 3 7) are different than most other fast sequences (typically starting with 1 4 9). This difference may imply that while sequences beginning with (1 4 9) may be very good at minimizing number of comparisons, it does not guarantee that they have a good overall running time.

\section{Conclusion}
Improvements for Shellsort traditionally come from finding gap sequences with better theoretical properties. Here we introduced an experimental framework to find improved gap sequences, following in the footsteps of \cite{ciura2001best}. Our generated gap sequences outperformed all well-known gap sequences in terms of number of comparisons on prescribed array sizes. Furthermore, the sequence Ours-A1000-Time is, to our knowledge, the function-based sequence with the quickest running time. However, it meets the performance of the Ciura sequence but does not surpass it. This may be improved with different sequence-generating functions or experimental setup, which we leave to future work. While the sequences presented here were optimized for chosen array sizes, the optimization may be repeated for any array size of interest.

{\small
\bibliography{optimization}
}

\end{document}